\input{aipcheck}

\documentclass[
    final            
  ]{aipproc}

\layoutstyle{6x9}

\begin{document}

\title{Symmetric Textures in SO(10) and LMA Solution for Solar Neutrinos}

\author{Mu-Chun Chen}{
  address={HET Group, Physics Department, Brookhaven National Laboratory,
Upton, NY 11973}
}

\author{K.T. Mahanthappa}{
  address={Department of Physics, University of Colorado, Boulder, CO 80309}
}

\begin{abstract}
{\small A model based on SUSY SO(10) combined with SU(2) family symmetry is 
constructed. In contrast with the commonly used effective operator approach, 
126-dimensional Higgs fields are utilized to construct the Yukawa sector. 
R-parity symmetry is thus preserved at low energies. The symmetric mass textures 
arising from the left-right symmetry breaking chain of SO(10) give rise to very 
good predictions for quark and lepton masses and mixings. The prediction for 
$\sin 2\beta$ agrees with the average of current bounds from BaBar and Belle. 
In the neutrino sector, our predictions are in good agreement with results
from atmospheric neutrino experiments. Our model accommodates the LMA 
solution to the solar neutrino anomaly. The
prediction of our model for the $|U_{e\nu_3}|$ element in the 
MNS matrix is close to the sensitivity of current experiments;
thus the validity of our model can be tested in the near future. We also 
investigate the correlation between the $|U_{e\nu_3}|$ 
element and $\tan^{2}\theta_{\odot}$ in a general two-zero 
neutrino mass texture. }
\end{abstract}

\maketitle



SO(10) has long been thought to be an attractive candidate for a
grand unified theory (GUT) for a number of reasons: First of all, it unifies
all the 15 known fermions with the right-handed neutrino for each family
into one 16-dimensional spinor representation. The seesaw mechanism then
arises very naturally, and the non-zero neutrino masses can thus be explained.
Since a complete quark-lepton symmetry is achieved, it has the promise for
explaining the pattern of fermion masses and mixing. Because B-L contained
in SO(10) is broken in symmetry breaking chain to the SM, it also has the
promise for baryogenesis. Recent atmospheric neutrino oscillation data from
Super-Kamiokande indicates non-zero neutrino masses. This in turn gives very
strong support to the viability of SO(10) as a GUT group. Models based on
SO(10) combined with discrete or continuous family symmetry have been
constructed to understand the flavor problem. Most of the models utilize 
``lopsided'' mass textures which usually require
more parameters and therefore are less constrained. Furthermore, the
right-handed neutrino Majorana mass operators in most of these models are made
out of $\scriptstyle 16_{H} \times 16_{H}$ which breaks the R-parity at a very
high scale. The aim of this talk, based on 
Ref.\cite{Chen:2000fp,Chen:2001pr,Chen:2002pa}, 
is to present a realistic model based on supersymmetric SO(10) combined
with SU(2) family symmetry which successfully predicts the low energy
fermion masses and mixings. Since we utilize {\it symmetric} mass textures and
$\scriptstyle \overline{126}$-dimensional Higgs representations for the right-handed
neutrino Majorana mass operator, our model is more constrained in addition to
having R-parity conserved. We first discuss the viable phenomenology of mass
textures followed by the model which accounts for it, and then the implications
of the model for neutrino mixing, CP violation, and neutrinoless double beta decay 
are presented.

The set of up- and down-quark mass matrix combination is given by,
at the GUT scale,
\begin{equation}
\scriptstyle{
\label{eq:Mud}
M_{u}=\left(
\begin{array}{ccc}
0 & 0 & a e^{i\gamma_{a}} \\
0 & b e^{i\gamma_{b}} & c e^{i\gamma_{c}} \\
a e^{i\gamma_{a}} & c e^{i\gamma_{c}} & e^{i\gamma_{d}}
\end{array}
\right) d v_{u}, \qquad 
M_{d}=\left(
\begin{array}{ccc}
0 & e e^{i\gamma_{e}} & 0 \\
e e^{i\gamma_{e}} & f e^{i\gamma_{f}} & 0 \\
0 & 0 & e^{i\gamma_h}
\end{array}
\right) h v_{d}}
\end{equation}
with $\scriptstyle a \simeq b \ll c \ll 1$, and $\scriptstyle e \ll f \ll 1$.
Symmetric mass textures arise naturally if SO(10) breaks down to the SM
through the left-right symmetric breaking chain $\scriptstyle SU(4) \times SU(2)_{L}
\times SU(2)_{R}$. SO(10) relates the up-quark mass matrix to 
the Dirac neutrino mass matrix, and the down-quark mass matrix to 
the charged lepton mass matrix. To achieve the Georgi-Jarlskog relations,  
a factor of -3 is needed
in the (2,2) entry of the charged lepton mass matrix,  
\begin{equation}\scriptstyle{
\label{eq:Me}
M_{e} = \left(
\begin{array}{ccc}
0 & e e^{i\gamma_{e}} & 0 \\
e e^{i\gamma_{e}} & -3 f e^{i\gamma_{f}} & 0 \\
0 & 0 & e^{i\gamma_h}
\end{array}
\right) h v_{d}}
\end{equation}
This factor of -3 can be accounted for by the SO(10) CG 
coefficients associated with $\scriptstyle \overline{126}$-dimensional
Higgs representations.  
The smallness of the neutrino masses is explained by 
the type I seesaw mechanism. 
The Dirac neutrino mass matrix is identical to
the mass matrix of the up-quarks in the framework of SO(10)   
\begin{equation}\scriptstyle{
\label{eq:Mlr}
M_{\nu_{LR}} = \left(
\begin{array}{ccc}
0 & 0 & a e^{i\gamma_{a}} \\
0 & b e^{i\gamma_{b}} & c e^{i\gamma_{c}} \\
a e^{i\gamma_{a}} & c e^{i\gamma_{c}} & e^{i\gamma_{d}}
\end{array}
\right) d v_{u}}
\end{equation}
The right-handed neutrino sector is an unknown sector. It is only constrained
by the requirement that it gives rise to a bi-maximal mixing pattern and a
hierarchical mass spectrum at low energies. To achieve this, we consider an
effective neutrino mass matrix of the form
\begin{equation}\scriptstyle{
\label{eq:Mll}
M_{\nu_{LL}}=M_{\nu_{LR}}^{T} M_{\nu_{RR}}^{-1} M_{\nu_{LR}}
= \left( 
\begin{array}{ccc}  
\scriptstyle{0} & \scriptstyle{0} & \scriptstyle{t} \\
\scriptstyle{0} & \scriptstyle{1} & \scriptstyle{1+t^{n}} \\  
\scriptstyle{t} & \scriptstyle{1+t^{n}} & \scriptstyle{1} 
\end{array} \right) \frac{d^{2}v_{u}^{2}}{M_{R}}} 
\end{equation}
The effective neutrino mass matrix of this form is obtained 
if the right-handed neutrino mass matrix has the same texture as that of the
Dirac  neutrino mass matrix, 
\begin{equation}\scriptstyle{
\label{eq:Mrr}
M_{\nu_{RR}}=
\left( \begin{array}{ccc}
\scriptstyle{0} & \scriptstyle{0} & \scriptstyle{\delta_{1}} \\
\scriptstyle{0} & \scriptstyle{\delta_{2}} & \scriptstyle{\delta_{3}} \\
\scriptstyle{\delta_{1}} & \scriptstyle{\delta_{3}} & \scriptstyle{1}
\end{array} \right) M_{R}}
\end{equation}
and if the elements $\scriptstyle \delta_{i}$ are of the right orders of magnitudes, 
determined by $\scriptstyle \delta_{i}=f_{i}(a,b,c,t,\theta)$, where 
$\scriptstyle \theta \equiv (\gamma_{b}-2\gamma_{c}-\gamma_{d})$.
Note that $\scriptstyle M_{\nu_{LL}}$ has the same texture as that of 
$\scriptstyle M_{\nu_{LR}}$ and
$\scriptstyle M_{\nu_{RR}}$, thus the seesaw mechanism is form 
invariant.  A generic feature of mass matrices of the type 
given in Eq.(\ref{eq:Mll}) is that they give rise to 
bi-maximal mixing pattern. After diagonalizing this mass matrix, one can see
immediately that the squared mass difference between $\scriptstyle m_{\nu_{1}}^{2}$ 
and $\scriptstyle m_{\nu_{2}}^{2}$ is of the order of
$\scriptstyle O(t^{3})$, while the squared mass difference between 
$\scriptstyle m_{\nu_{2}}^{2}$ and $\scriptstyle m_{\nu_{3}}^{2}$ 
is of the order of $\scriptstyle O(1)$, in units of $\scriptstyle \Lambda$. 
For $\scriptstyle t \ll 1$, the phenomenologically favored relation 
$\scriptstyle \Delta m_{atm}^{2} \gg \Delta m_{\odot}^{2}$ is thus obtained.

The SU(2) family symmetry is implemented {\it {\'a} la} the
Froggatt-Nielsen mechanism. The heaviness of the top quark and to suppress the
SUSY FCNC together suggest that the third family of matter fields transform
as a singlet and the lighter two families of matter fields transform as a
doublet under SU(2). In the family symmetric limit, only the third family has
non-vanishing Yukawa couplings. SU(2) breaks down in two steps:
$\scriptstyle SU(2) \stackrel{\epsilon M}{\longrightarrow} 
U(1) \stackrel{\epsilon' M}{\longrightarrow}
nothing$, 
where $\scriptstyle \epsilon' \ll \epsilon \ll 1$ and $\scriptstyle M$ 
is the family symmetry scale. 
These small parameters $\scriptstyle \epsilon$ and 
$\scriptstyle \epsilon'$ are the ratios of 
the vacuum expectation
values of the flavon fields to the family symmetry scale. 
A discrete symmetry $\scriptstyle (Z_{2})^{3}$ is needed to avoid unwanted couplings.
The field content of our model is then given by\\
-- matter fields
\begin{displaymath}\scriptstyle{
\psi_{a} \sim (16,2)^{-++} \quad (a=1,2), \qquad
\psi_{3} \sim (16,1)^{+++}}
\end{displaymath}
-- Higgs fields:
\begin{displaymath}
\begin{array}{ll}
\scriptstyle{
(10,1):} & \scriptstyle{\quad T_{1}^{+++}, \quad T_{2}^{-+-},\quad
T_{3}^{--+}, \quad T_{4}^{---}, \quad T_{5}^{+--}}
\\
\scriptstyle{
(\overline{126},1):} & \scriptstyle{\quad \overline{C}^{---}, 
\quad \overline{C}_{1}^{+++},
\quad \overline{C}_{2}^{++-}}
\end{array}
\end{displaymath}
-- Flavon fields:
\begin{displaymath}
\begin{array}{ll}
\scriptstyle{(1,2): } 
& 
\scriptstyle{\quad \phi_{(1)}^{++-}, \quad \phi_{(2)}^{+-+}, \quad \Phi^{-+-}}
\\
\scriptstyle{(1,3):} 
&
\scriptstyle{\quad S_{(1)}^{+--}, \quad S_{(2)}^{---}, \quad
\Sigma^{++-}}
\end{array}
\end{displaymath}
and the superpotential of our model which generates fermion masses is given by
\begin{equation}\scriptstyle{
\begin{array}{l}
\scriptstyle{W = W_{D(irac)} + W_{M(ajorana)}}
\\
\scriptstyle{W_{D}=\psi_{3}\psi_{3} T_{1} + \frac{1}{M} \psi_{3} \psi_{a}
\left(T_{2}\phi_{(1)}
+T_{3}\phi_{(2)}\right)
+ \frac{1}{M} \psi_{a} \psi_{b} \left(T_{4} + \overline{C}\right) S_{(2)}
+ \frac{1}{M} \psi_{a} \psi_{b} T_{5} S_{(1)}}
\\
\scriptstyle{W_{M}=\psi_{3} \psi_{3} \overline{C}_{1} 
+ \frac{1}{M} \psi_{3} \psi_{a} \Phi \overline{C}_{2}
+ \frac{1}{M} \psi_{a} \psi_{b} \Sigma \overline{C}_{2}}
\end{array}}
\end{equation}
where $\scriptstyle T_{i}$'s and $\scriptstyle \overline{C}_{i}$'s 
are the $\scriptstyle 10$ and $\scriptstyle \overline{126}$
dimensional Higgs representations of SO(10) respectively, and $\scriptstyle \Phi$ and 
$\scriptstyle \Sigma$ are the 
doublet and triplet of SU(2), respectively. 
Detailed quantum number assignment
and the VEVs acquired by various scalar fields are given in Ref.\cite{Chen:2000fp}. 
This superpotential gives rise to the mass textures given in 
Eq.(\ref{eq:Mud})-(\ref{eq:Mrr}):
\begin{equation}\scriptstyle{
M_{u,\nu_{LR}}=
\left( \begin{array}{ccc}
\scriptstyle{0} & 
\scriptstyle{0} & 
\scriptstyle{\left<10_{2}^{+} \right> \epsilon'}\\
\scriptstyle{0} & 
\scriptstyle{\left<10_{4}^{+} \right> \epsilon} & 
\scriptstyle{\left<10_{3}^{+} \right> \epsilon} \\
\scriptstyle{\left<10_{2}^{+} \right> \epsilon'} & \
\scriptstyle{\left<10_{3}^{+} \right> \epsilon} &
\scriptstyle{\left<10_{1}^{+} \right>}
\end{array} \right)
= 
\left( \begin{array}{ccc}
\scriptstyle{0} & 
\scriptstyle{0} & 
\scriptstyle{r_{2} \epsilon'}\\
\scriptstyle{0} & 
\scriptstyle{r_{4} \epsilon} & 
\scriptstyle{\epsilon} \\
\scriptstyle{r_{2} \epsilon'} & 
\scriptstyle{\epsilon} & 
\scriptstyle{1}
\end{array} \right) M_{U}}
\end{equation}
\begin{equation}\scriptstyle{
M_{d,e}=
\left(\begin{array}{ccc}
\scriptstyle{0} & 
\scriptstyle{\left<10_{5}^{-} \right> \epsilon'} & 
\scriptstyle{0} \\
\scriptstyle{\left<10_{5}^{-} \right> \epsilon'} &  
\scriptstyle{(1,-3)\left<\overline{126}^{-} \right> \epsilon} & 
\scriptstyle{0}\\ 
\scriptstyle{0} & 
\scriptstyle{0} & 
\scriptstyle{\left<10_{1}^{-} \right>}
\end{array} \right)
=
\left(\begin{array}{ccc}
\scriptstyle{0} & 
\scriptstyle{\epsilon'} & 
\scriptstyle{0} \\
\scriptstyle{\epsilon'} &  
\scriptstyle{(1,-3) p \epsilon} & 
\scriptstyle{0}\\
\scriptstyle{0} & 
\scriptstyle{0} & 
\scriptstyle{1}
\end{array} \right) M_{D}}
\end{equation}
where
$\scriptstyle M_{U} \equiv \left<10_{1}^{+} \right>$, 
$\scriptstyle M_{D} \equiv \left<10_{1}^{-} \right>$, 
$\scriptstyle r_{2} \equiv \left<10_{2}^{+} \right> / \left<10_{1}^{+} \right>$, 
$\scriptstyle r_{4} \equiv \left<10_{4}^{+} \right> / \left<10_{1}^{+} \right>$ and
$\scriptstyle p \equiv \left<\overline{126}^{-}\right> / \left<10_{1}^{-} \right>$.
The right-handed neutrino mass matrix is  
\begin{equation}\scriptstyle{
M_{\nu_{RR}}=  
\left( \begin{array}{ccc}
\scriptstyle{0} & 
\scriptstyle{0} & 
\scriptstyle{\left<\overline{126}_{2}^{'0} \right> \delta_{1}}\\
\scriptstyle{0} & 
\scriptstyle{\left<\overline{126}_{2}^{'0} \right> \delta_{2}} & 
\scriptstyle{\left<\overline{126}_{2}^{'0} \right> \delta_{3}} \\ 
\scriptstyle{\left<\overline{126}_{2}^{'0} \right> \delta_{1}} & 
\scriptstyle{\left<\overline{126}_{2}^{'0} \right> \delta_{3}} &
\scriptstyle{\left<\overline{126}_{1}^{'0} \right> }
\end{array} \right)
= 
\left( \begin{array}{ccc}
\scriptstyle{0} & 
\scriptstyle{0} & 
\scriptstyle{\delta_{1}}\\
\scriptstyle{0} & 
\scriptstyle{\delta_{2}} & 
\scriptstyle{\delta_{3}} \\ 
\scriptstyle{\delta_{1}} & 
\scriptstyle{\delta_{3}} & 
\scriptstyle{1}
\end{array} \right) M_{R}
\label{Mrr}}
\end{equation}
with $\scriptstyle M_{R} \equiv \left<\overline{126}^{'0}_{1}\right>$.
Note that, since we use
$\overline{126}$-dimensional representations of Higgses to generate the heavy
Majorana neutrino mass terms, R-parity is preserved at all energies. 

With values of $\scriptstyle m_{f}, (f= u,c,t,e,\mu,\tau)$ and those of 
$\scriptstyle |V_{us, ub, cb}|$
at the weak scale, the input parameters at the GUT scale are determined. The
predictions for the charged fermion masses and CKM mixing 
of our model at $\scriptstyle M_{Z}$ which are summarized
below including 2-loop RGE effects are in good agreements
with the experimental values:
\begin{equation}
\begin{array}{lllll}
\scriptstyle{\frac{m_{s}}{m_{d}}=25}, &
\scriptstyle{m_{s} = 85.66 MeV}, &
\scriptstyle{m_{b} = 3.147 GeV}, &
\scriptstyle{\vert V_{ud} \vert = 0.9751}, &
\scriptstyle{\vert V_{cd} \vert = 0.2218} \\
\scriptstyle{\vert V_{cs} \vert = 0.9744}, &
\scriptstyle{\vert V_{td} \vert = 0.005358}, &
\scriptstyle{\vert V_{ts} \vert = 0.03611}, &
\scriptstyle{\vert V_{tb} \vert = 0.9993}, &
\scriptstyle{J_{CP}^{q} = 1.748 \times 10^{-5}} \\
\scriptstyle{\sin 2\alpha = -0.8913}, &
\scriptstyle{\sin 2\beta = 0.7416}, &
\scriptstyle{\gamma = 34.55^{0}}. &&
\end{array}
\end{equation}
Using the mass square differences  
$\scriptstyle{\Delta m_{atm}^{2}=2.78 \times 10^{-3} \; eV^{2}}$ and  
$\scriptstyle{\Delta m_{\odot}^{2}=7.25 \times 10^{-5} \; eV^{2}}$ for the LOW 
solution as input parameters, we determine 
$\scriptstyle{(t, M_{R}) = (0.35,5.94 \times 10^{12} GeV)}$, and correspondingly  
$\scriptstyle{(\delta_{1},\delta_{2},\delta_{3}) 
= (0.00119,0.000841 e^{i \; (0.220)},0.0211 e^{-i \;(0.029)})}$. 
The three mass eigenvalues are predicted to be  
$\scriptstyle{
(m_{\nu_{1}},m_{\nu_{2}},m_{\nu_{3}}) = (0.00363,0.00926,0.0535) \; eV}$, 
and the mixing angles are predicted to be
$\scriptstyle{\sin^{2} 2 \theta_{atm} = 1, \;
\tan^{2} \theta_{\odot} = 0.58, \;
\sin^{2}\theta_{13} = 0.022}$. These predictions agree with current 
bounds from experiments within $1 \; \sigma$. 
The strengths of CP violation in the lepton sector are
$\scriptstyle{(J_{CP}^{l},\alpha_{31},\alpha_{21}) = (-0.00690,0.490,-2.29)}$, 
and the matrix element for the neutrinoless double 
$\beta$ decay is given by   
$\scriptstyle{\vert < m > \vert = 2.22 \times 10^{-3} \; eV}$.
The correlation between 
$\scriptstyle{|U_{e\nu_{3}}|^{2}}$ and 
$\scriptstyle{\tan^{2}\theta_{\odot}}$ is plotted 
in Fig. 1.
\begin{figure}
  \includegraphics[height=.23\textheight]{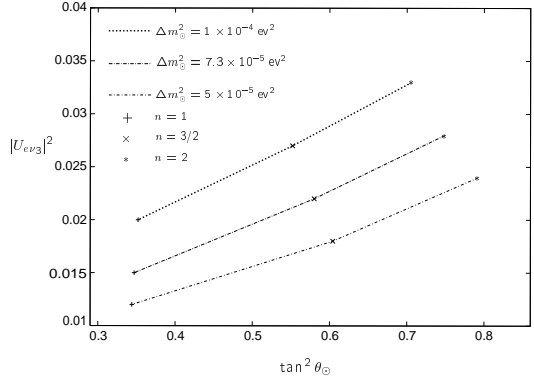}
\caption{\label{corelate}Correlation between $|U_{e\nu_{3}}|^{2}$ and 
$\tan^{2}\theta_{\odot}$. $\Delta m_{atm}^{2}$ is $2.8 \times 10^{-3} eV^{3}$. 
The dotted line corresponds to the upper bound 
$\Delta m_{\odot}^{2} = 10^{-4} eV^{2}$; the dotted-long-dashed 
line corresponds to the best fit value $\Delta m_{\odot}^{2} 
= 7.3 \times 10^{-5} eV^{2}$; the dotted-short-dashed 
line corresponds to the lower bound $\Delta m_{\odot}^{2} 
= 5 \times 10^{-5} eV^{2}$.}
\end{figure}
A comparison of the predictions for $\scriptstyle{\sin^{2}2\theta_{13}}$ 
and $\scriptstyle{(\alpha,\beta,\gamma)}$ from different SO(10) 
models is given in Ref.[4]. This work was supported by US DOE under Grant No. 
DE-AC02-98CH10886 and DE-FG03-95ER40894.



\bibliographystyle{aipproc}   

\begin{thebibliography}{99}

\small{

\bibitem{Chen:2000fp}
{\small M.~-C.~Chen and K.~T.~Mahanthappa,
{\it Phys. Rev.} {\bf D62}, 113007 (2000).}


\bibitem{Chen:2001pr}
{\small
M.~-C.~Chen and K.~T.~Mahanthappa,
{\it Phys. Rev.} {\bf D65}, 053010 (2002).}


\bibitem{Chen:2002pa}
\small{
M.~-C.~Chen and K.~T.~Mahanthappa,
{\it Phys. Rev.} {\bf D68}, 017301 (2003). }


\bibitem{Chen:2003zv}
\small{
M.~-C.~Chen and K.~T.~Mahanthappa,
arXiv:hep-ph/0305088.}

}
\end{thebibliography}

\end{document}